\documentclass[paper,notoc,11pt]{JHEP3}

\usepackage{graphicx}
\usepackage{cite}

\usepackage{pslatex}
\usepackage[latin1]{inputenc}
\usepackage[T1]{fontenc}

\def\z#1{{\zeta_{#1}}}
\def\zs{{\zeta_{2}^{\,2}}}
\def\zt{{\zeta_{2}^{\,3}}}
\def\zf{{\zeta_{2}^{\,4}}}
\def\zzs{{\zeta_{3}^{\,2}}}
\def\ca{{C^{}_A}}
\def\cf{{C^{}_F}}
\def\nf{{n^{}_{\! f}}}
\def\nsq{{n^{\,2}_{\! f}}}
\def\cas{{C^{\,2}_A}}
\def\cf{{C^{}_F}}
\def\cfs{{C^{\, 2}_F}}
\def\cft{{C^{\, 3}_F}}

\newcommand{\beq}{\begin{equation}}
\newcommand{\eeq}{\end{equation}}
\newcommand{\bea}{\begin{eqnarray}}
\newcommand{\eea}{\end{eqnarray}}
\newcommand{\nn}{\nonumber}
\newcommand{\ra}{\rightarrow}
\newcommand\MSb{$\overline{\mbox{MS}}$}

\newcommand{\ep}{\epsilon}
\newcommand{\as}{\alpha_{\rm s}}
\newcommand{\ars}{a_{\rm s}}
\newcommand{\DD}{{\cal D}}

\author{
S. Moch$^{\, a}$, J.A.M. Vermaseren$^{\, b}$ and A. Vogt$^{\, c}$\\
$^a$Deutsches Elektronensynchrotron DESY, Platanenallee 6, D--15738 Zeuthen, 
Germany \\
$^b$NIKHEF Theory Group, Kruislaan 409, 1098 SJ Amsterdam, The Netherlands \\
$^c$IPPP, Department of Physics, Durham University, South Road, Durham DH1 3LE,
UK}

\title{The Quark Form Factor at Higher Orders}

\preprint{hep-ph/0507039 \\[2mm] DESY 05-106, SFB/CPP-05-26 
\\ DCPT/05/66, IPPP/05/33 \\ NIKHEF 05-011 \\$\,$}

\abstract{
We study the electromagnetic on-shell form factor of quarks in massless 
perturbative QCD. We derive the complete pole part in dimensional 
regularization at three loops, and extend the resummation of the form factor 
to the next-to-next-to-leading contributions. These results are employed to 
evaluate the infrared finite absolute ratio of the time-like and space-like
form factors up to the fourth order in the strong coupling constant. Besides
for the pole structure of higher-loop QCD amplitudes, our new contributions to
the form factor are also relevant for the high-energy limit of massive gauge
theories like QED. The highest-transcendentality component of our results 
confirms a result recently obtained in $\,{\cal N}\! = 4\,$ Super-Yang-Mills 
theory.  
}

\keywords{QCD, multi-loop computations, electromagnetic processes and 
properties, \\ \hspace*{2.1cm}extended supersymmetry}

\begin{document}
\section{Introduction}
\label{sec:intro}
The electromagnetic form factor of quarks is a quantity of considerable 
interest in Quantum Chromodynamics (QCD) and in gauge theories in general.
At high photon virtualities $Q^2$ this quantity receives double logarithmic
corrections of infrared and collinear origin~\cite{Sudakov:1954sw}, which take
the form of double poles in dimensional regularization for the case of 
massless on-shell quarks studied in the present article. 
These contributions can be resummed by evolution equations in $Q^2$ based on 
universal factorization properties of the amplitude in the relevant kinematic 
limit, resulting in the well-known exponentiation of the form factor
\cite{Mueller:1979ih,Collins:1980ih,Sen:1981sd}.
So far perturbative calculations have been performed up to two loops for
both the massless on-shell case~\cite{vanNeerven:1985xr,Matsuura:1989sm} and 
heavy quarks~\cite{Bernreuther:2004ih}. Accordingly, the exponentiation has 
been studied up to the next-to-leading (NL) contributions
\cite{Korchemsky:1988hd,Magnea:1990zb,Magnea:2000ss}.

Higher-order corrections to the quark form factor are not only of general
interest in quantum field theory, but also relevant for practical applications,
as this quantity contributes to phenomenologically important processes. 
Research in the past years has yielded dramatic progress in next-to-next-to-%
leading order (NNLO) perturbative calculations, see, for example, Ref.~\cite
{Gehrmann-DeRidder:2005cm} and numerous references therein. 
This progress also led to further investigations of the general structure 
of amplitudes and cross section at higher loop-orders, which in turn further 
stimulated the interest in all-order resummations. Consequently, the intimate 
connection between resummation and perturbative results at multiple loops has 
become much more prominent~\cite{Catani:1998bh,Sterman:2002qn,Kosower:2003cz}.

Very recently, we have presented the first complete calculation of the third-%
order corrections to a hard-scattering observable depending on a dimensionless
variable, the structure function $F_2$ in photon-exchange deep-inelastic
scattering~\cite{Vermaseren:2005qc}. After exploring the consequences of that
result for the soft-gluon threshold resummation in Ref.~\cite{Moch:2005ba},
we here present its implications on the quark form factor (from now on always
referring to the massless on-shell case, if not explicitly indicated 
otherwise) and its resummation. We are able, after extending the two-loop form
factor beyond the previous order $\ep^0$ in dimensional regularization, to 
derive the complete series of poles, $\ep^{-6}\:\ldots\:\ep^{-1}$, at three 
loops.  These terms in turn provide the coefficients required to extend the 
exponentiation of the form factor to the next-to-next-to-leading (NNL)
contributions which we work out explicitly.   

This article is organized as follows: 
In Section 2 we address the resummation of the quark form factor. We briefly
recall the evolution equation and its solution, and present the explicit
expansion up to four loops in terms of two perturbative functions, $A(\as)$
(known up to three loops from the NNLO splitting functions~\cite
{Moch:2004pa,Vogt:2004mw}) and $G(\as,\ep)$. 
In Section 3 we extend the two-loop form factor to order $\ep^2$ and extract 
the pole terms at three loops from our structure-function calculation
\cite{Vermaseren:2005qc}. These results are employed to extend the first- and
second-order parts of the resummation function $G$ to higher orders in $\ep$,
and to derive the leading-$\ep$ term at the third order in the strong coupling.
Some first implications of these results are discussed in Section 4. 
Here we extend the ratio of the time-like and space-like form factors
\cite{Magnea:1990zb} to the fourth order in $\as$, compare to a recent result 
for ${\cal N}= 4$ Super-Yang-Mills theory~\cite{Bern:2005iz} and indicate 
applications on the infrared structure of massive gauge theories
\cite{Kuhn:2001hz,Jantzen:2005xi}.
We briefly summarized our results in Section 5. 
A few technical details for the solution of the evolution equations in 
Section 2 can be found in Appendix A. Finally the break-up of the 
($\ep$-extended) two-loop form factor into its Feynman diagrams is presented in 
Appendix~B. 
\section{The resummation of the quark form factor}
\label{sec:ffactor2}
The subject of our study are the QCD corrections to the $\gamma^{\,\ast}\!qq$ 
(or $\gamma^{\,\ast}\!q\bar{q}$) vertex, where $\gamma^{\,\ast}$ denotes a 
space-like (or time-like) photon with virtuality $Q^2$, and $q/\bar{q}$ a 
massless external quark$\,$/$\,$antiquark. Until Section 4 we will focus on the 
space-like case, thus the relevant amplitude is
\beq
\label{eq:fmudeq}
\Gamma_\mu  \: = \: {\rm i} e_{\rm q}\,
\bigl({\bar u}\, \gamma_{\mu\,} u\bigr)\, {\cal F}\! (\as,Q^2)\, ,
\eeq
where the scalar function ${\cal F}$ on the right-hand side is the 
space-like quark form 
factor. This quantity can be calculated order by order in the strong coupling 
constant $\as$ and, as mentioned above, is so far known to two loops
\cite{vanNeerven:1985xr,Matsuura:1989sm}. ${\cal F}$ is gauge invariant, but 
divergent. As usual we work in dimensional regularization with $D=4-2\epsilon$,
thus these divergences show up as poles $\ep^{-k}$ in the present article.

The exponentiation of the form factor, which extends beyond the resummation of 
renormalization group logarithms, is achieved by solving the well-known 
evolution equations~\cite
{Collins:1980ih,Korchemsky:1988hd,Sen:1981sd,Magnea:1990zb,Magnea:2000ss}
\beq
\label{eq:ffdeq}
Q^2 {\partial \over \partial Q^2} \ln {\cal F}\!\left(\as, {Q^2 \over \mu^2}, 
\epsilon\right) \:  = \:  
  {1 \over 2} \: K(\as,\epsilon) 
+ {1 \over 2} \: G\left({Q^2 \over \mu^2},\as,\epsilon \right) \, .
\eeq
Here $\mu$ represents the renormalization scale, and the functions $G$ and 
$K$ are subject to the renormalization group equations~\cite{Collins:1980ih}
\bea
\label{eq:Gdeq}
\left(\mu^2 {\partial \over \partial \mu^2} + \beta(\as,\epsilon)\, 
{\partial \over \partial \as} \right) 
G\left({Q^2 \over \mu^2},\as,\epsilon \right) & = & ~~~ A(\as) \, , \\[1mm]
\label{eq:Kdeq}
\left(\mu^2 {\partial \over \partial \mu^2} + \beta(\as,\epsilon)\, 
{\partial \over \partial \as} \right) K(\as,\epsilon) & = & - A(\as) \, .
\eea
All infrared singularities are collected by the scale-independent function $K$,
which in the \MSb\ scheme consists of a series of poles in $\epsilon$. The 
function $G$, on the other hand, is finite for $\epsilon \ra 0$ and includes 
all dependence on the scale $Q^2$.
The renormalization properties of $G$ and $K$ are both governed by the same 
anomalous dimension $A$, because the sum of $G$ and $K$ is a renormalization 
group invariant. This quantity is given by a power expansion in the strong 
coupling, for which we use the convention (also employed for all other 
expansions in $\as$ throughout this article)
\beq
\label{eq:abexp}
  A(\as) 
\: = \: \sum\limits_{i=1}^{\infty} \left({\as \over 4\pi}\right)^i\, A_i
\: \equiv \: \sum\limits_{i=1}^{\infty} \ars^{\,i}\,  A_i\, . 
\eeq
In fact, the anomalous dimension $A$ also occurs in many other circumstances, 
for instance as the coefficient of the $1/(1-x)_+$ contribution to the 
Altarelli-Parisi quark-quark splitting function and as the anomalous dimension 
of a Wilson line with a cusp~\cite{Korchemsky:1989si}.

As already indicated by the argument of the beta function, the solution of 
Eqs.~(\ref{eq:Gdeq}) and (\ref{eq:Kdeq}) requires the running coupling in 
$D$ dimensions. Following Refs.~\cite{Magnea:2000ss,Contopanagos:1997nh} we 
define ${\bar a}(\lambda,a_s,\epsilon)$, where $\lambda$ is a dimensionless 
ratio of scales like $\lambda = Q^2/\mu^2$. The resummation of the NNL 
contributions to the form factor requires the scale dependence of ${\bar a}$ to 
NNLO accuracy~\cite{Tarasov:1980au,Larin:1993tp} (see the discussion at the
end of this section), obtained by solving
\bea
\label{eq:betaD}
  \lambda {\partial \over \partial \lambda} {\bar a}(\lambda,\ars,\epsilon) &=& 
- \epsilon\, {\bar a}(\lambda,\ars,\epsilon)\
- \beta_0\,  {\bar a}^2(\lambda,\ars,\epsilon) 
- \beta_1\,  {\bar a}^3(\lambda,\ars,\epsilon) 
- \beta_2\,  {\bar a}^4(\lambda,\ars,\epsilon)  
\eea
with the boundary condition ${\bar a}(1,\ars,\epsilon) = \ars$. Extending the
result of Ref.~\cite{Contopanagos:1997nh} by one order, this solution is
given by
\bea
\label{eq:arunD}
{\bar a}(\lambda,\ars,\epsilon) &=&
    {\ars \over X} \biggl\{ 1 - \epsilon {\beta_1 \over \beta_0^2} 
    {\ln X \over X} \biggr\}   
  - {\ars^{\,2} \over X^2} \biggl\{ {\beta_1 \over \beta_0} 
    (\ln X + Y) \biggr\}
  + {\ars^{\,3} \over X^3} \biggl\{
    {\beta_1^2 \over \beta_0^2} {3 \over 2} \ln^2 X 
    \left( 1 + Y + {1 \over 4} Y^2 \right) \quad
\nonumber \\[1mm]
& & \mbox{}
    + {\beta_2 \over \beta_0} \ln X 
      \left( {1 \over 6} (3 + Y) (1-X) - 1 - Y 
      - {1 \over 3} Y^2 \right)
    \biggr\}
  + {\cal O} \left( \ars^{\,4} \right)\, ,
\eea
where we have used the abbreviations
\beq
\label{eq:XYparD}
  X \:=\: 1 - \ars {\beta_0 \over \epsilon} (\lambda^{-\epsilon} - 1)
  \:\: , \quad\quad
  Y \: = \: {\epsilon (1-X) \over \ars \beta_0} \, .
\eeq
With Eq.~(\ref{eq:arunD}) for the running coupling, Eq.~(\ref{eq:Gdeq}) can 
now be solved to the required accuracy,
\bea
\label{eq:Gsol}
G\left({Q^2 \over \mu^2},\as,\epsilon \right) & = & 
G\left(1,{\bar a}\left({Q^2 \over \mu^2},\ars,\epsilon \right),\epsilon \right) 
+ \int\limits_{Q^2/\mu^2}^1\, {d \lambda \over \lambda}\
A({\bar a}(\lambda,\ars,\epsilon))\, .
\eea
The perturbative expansion of the boundary condition $G(1,{\bar a},\epsilon)$ 
can be derived by comparison to the fixed-order results for the form factor.

After recursively determining (see, e.g., Ref.~\cite{Magnea:2000ss} for 
details) the scale-independent counter-term function $K$ from 
Eq.~(\ref{eq:Kdeq}), the resummed quark form factor reads
\bea
\label{eq:resummedff}
\ln {\cal F}\!\left(\as, {Q^2 \over \mu^2}, \epsilon\right) & = & 
{1 \over 2} \int\limits_0^{Q^2/\mu^2}\, {d \xi \over \xi} \Bigg( 
  K(\as,\epsilon) + G(1,{\bar a}(\xi,a_s,\epsilon),\epsilon) 
  + \int\limits_\xi^1\, {d \lambda \over \lambda} 
  A({\bar a}(\lambda,\ars,\epsilon)) \Bigg) \quad
\eea
with the boundary condition ${\cal F}(\alpha_s,0,\epsilon) = 1$ 
\cite{Magnea:1990zb}.
After expanding the $D$-dimensional coupling according to Eq.~(\ref{eq:arunD}), 
$\ln {\cal F}$ exhibits double logarithms of $Q^2/\mu^2$ and double poles in 
$\epsilon$, which are generated by the two integrations. In addition the 
integral over the anomalous dimension $A$ leads to terms which are independent 
of the outer integration variable $\xi$.  These logarithmic singularities at 
$\xi=0$ are canceled by the function $K$ order by order in the perturbative 
expansion.
 
The well-known relation (\ref{eq:resummedff}) can be employed either for a 
direct evaluation of the form factor due to the analyticity in $D$ dimensions
\cite{Magnea:2000ss}
or, by means of finite-order expansions and matching, for predictions of 
perturbative results at higher orders. Here we will focus on the latter issue. 
In particular, we will derive explicit results at three and four loops.
This is done by performing the integrations in Eq.~(\ref{eq:resummedff}) after 
inserting the perturbative expansions of all quantities. The resulting 
integrals can be evaluated using algorithms for the evaluations of nested sums
\cite{Vermaseren:1998uu,Moch:2001zr}.
Some technical details for this step are given in Appendix~A, where 
Eqs.~(\ref{eq:auxint1})--(\ref{eq:auxint4}) represent sample types of relevant 
integrals. Further details may also be found in Ref.~\cite{Magnea:2000ss}.
 
It is convenient to express the loop-expanded form factor in 
terms of the bare (unrenormalized) coupling $\as^{\rm{b}}$ instead of
the renormalized coupling $\as$ as in Eq.~(\ref{eq:resummedff}). 
The couplings $\as^{\rm{b}}$ and $\as$ are related~by
\beq
\label{eq:alpha-s-renorm}
\as^{\rm{b}} \: = \:  Z_{\as}\, \as\, ,
\eeq
with the renormalization constant $Z_{\as}$ in the \MSb\ scheme given by
\beq
\label{eq:alpha-Z}
Z_{\as} \: = \: 1 - {\beta_0 \over \epsilon} \ars\, 
+ \left({\beta_0^2 \over \epsilon^2} 
  - {1 \over 2} {\beta_1 \over \epsilon}\right) \ars^{\,2}\,
- \left({\beta_0^3 \over \epsilon^3} 
   - {7 \over 6} {\beta_1 \beta_0 \over \epsilon^2} 
   + {1 \over 3} {\beta_2 \over \epsilon}\right) \ars^{\,3}\, ,
\eeq
and also the bare expansion parameter normalized as $\ars^{\rm{b}} = 
\as^{\rm{b}} / (4\pi)\,$.
The perturbative expansion of the bare (unrenormalized) quark form factor then 
reads
\beq
\label{eq:finiteorderff}
  {\cal F}^{\rm{b}}(\alpha_s^{\rm{b}},Q^2) \: = \: 
  1 + \sum\limits_{l=1}^\infty\, \bigl(a_s^{\rm{b}} \bigr)^l \, 
  \biggl({Q^2 \over \mu^2}\biggl)^{\! -l\epsilon}\, {\cal F}_l^{}\,  \, .
\eeq
In terms of the $i$-th order parameters $A_i$ in Eq.~(\ref{eq:abexp}) and the
corresponding functions $G_i(\ep)$, the expansion coefficients up to four
loops read
\bea
  {\cal F}_1^{} & = & 
          - {1 \over 2} \* {1 \over \epsilon^2} \* A_1
          - {1 \over 2} \* {1 \over \epsilon} \* G_1
\label{eq:ff1loop}
\, ,  \\[1mm]
  {\cal F}_2^{} & = & 
            {1 \over 8} \* {1 \over \epsilon^4} \* A_1^2
          + {1 \over 8} \* {1 \over \epsilon^3} \* A_1 \* ( 
            2 \* G_1
          - \beta_0
          )
          + {1 \over 8} \* {1 \over \epsilon^2} \* (
            G_1^2 
          - A_2 
          - 2 \* \beta_0 \* G_1
          )
          - {1 \over 4} \* {1 \over \epsilon} \* G_2
\label{eq:ff2loop}
\, ,  \\[1mm]
  {\cal F}_3^{} & = & 
          - {1 \over 48} \* {1 \over \epsilon^6} \* A_1^3
          - {1 \over 16} \* {1 \over \epsilon^5} \* A_1^2 \* ( 
            G_1 
          - \beta_0
          )
          - {1 \over 144} \* {1 \over \epsilon^4} \* A_1 \* (
            9 \* G_1^2 
          - 9 \* A_2 
          - 27 \* \beta_0 \* G_1 
          + 8 \* \beta_0^2
          )
\nonumber\\
& &\mbox{}
          - {1 \over 144} \* {1 \over \epsilon^3} \* ( 
            3 \* G_1^3 
          - 9 \* A_2 \* G_1 
          - 18 \* A_1 \* G_2 
          + 4 \* \beta_1 \* A_1 
          - 18 \* \beta_0 \* G_1^2 
          + 16 \* \beta_0 \* A_2 
          + 24 \* \beta_0^2 \* G_1
          )
\nonumber\\
& &\mbox{}
          + {1 \over 72} \* {1 \over \epsilon^2} \* (
            9 \* G_1 \* G_2 
          - 4 \* A_3 
          - 6 \* \beta_1 \* G_1 
          - 24 \* \beta_0 \* G_2
          )
          - {1 \over 6} \* {1 \over \epsilon} \* G_3
\label{eq:ff3loop}
\, ,  \\[1mm]
  {\cal F}_4^{} & = & 
            {1 \over 384} \* {1 \over \epsilon^8} \* A_1^4
          + {1 \over 192} \* {1 \over \epsilon^7} \* A_1^3 \* (
            2 \* G_1 
          - 3 \* \beta_0
          )
          + {1 \over 1152} \* {1 \over \epsilon^6} \* A_1^2 \* (
            18 \* G_1^2 
          - 18 \* A_2 
          - 72 \* \beta_0 \* G_1 
          + 41 \* \beta_0^2
          )
\nonumber\\
& &\mbox{}
          + {1 \over 576} \* {1 \over \epsilon^5} \* A_1 \* (
            6 \* G_1^3 
          - 18 \* A_2 \* G_1 
          - 18 \* A_1 \* G_2 
          + 8 \* \beta_1 \* A_1 
          - 45 \* \beta_0 \* G_1^2 
          + 41 \* \beta_0 \* A_2 
          + 82 \* \beta_0^2 \* G_1 
          - 18 \* \beta_0^3
          )
\nonumber\\
& &\mbox{}
          + {1 \over 1152} \* {1 \over \epsilon^4} \* (
            3 \* G_1^4 
          - 18 \* A_2 \* G_1^2 
          + 9 \* A_2^2 
          - 72 \* A_1 \* G_1 \* G_2 
          + 32 \* A_1 \* A_3
          + 64 \* \beta_1 \* A_1 \* G_1 
          - 36 \* \beta_0 \* G_1^3 
\nonumber\\
& &\mbox{}
          + 100 \* \beta_0 \* A_2 \* G_1 
          + 228 \* \beta_0 \* A_1 \* G_2 
          - 48 \* \beta_0 \* \beta_1 \* A_1 
          + 132 \* \beta_0^2 \* G_1^2 
          - 108 \* \beta_0^2 \* A_2 
          - 144 \* \beta_0^3 \* G_1
          )
\nonumber\\
& &\mbox{}
          + {1 \over 288} \* {1 \over \epsilon^3} \* ( 
          - 9 \* G_1^2 \* G_2 
          + 8 \* A_3 \* G_1 
          + 9 \* A_2 \* G_2 
          + 24 \* A_1 \* G_3 
          - 3 \* \beta_2 \* A_1 
          + 12 \* \beta_1 \* G_1^2 
          - 9 \* \beta_1 \* A_2 
\nonumber\\
& &\mbox{}
          + 66 \* \beta_0 \* G_1 \* G_2 
          - 27 \* \beta_0 \* A_3 
          - 48 \* \beta_0 \* \beta_1 \* G_1
          - 108 \* \beta_0^2 \* G_2
          )
          + {1 \over 96} \* {1 \over \epsilon^2} \* (
            3 \* G_2^2 
          + 8 \* G_1 \* G_3 
          - 3 \* A_4 
\nonumber\\
& &\mbox{}
          - 4 \* \beta_2 \* G_1 
          - 12 \* \beta_1 \* G_2 
          - 36 \* \beta_0 \* G_3
          )
          - {1 \over 8} \*  {1 \over \epsilon} \* G_4
\label{eq:ff4loop}
\, .
\eea
The three- and four-loop relations~(\ref{eq:ff3loop}) and (\ref{eq:ff4loop}) 
are new results of the present article.
Recall that Eqs.~(\ref{eq:ff1loop}) -- (\ref{eq:ff4loop}) directly refer to
the bare form factor. The corresponding renormalized results can be derived 
with the help of Eqs.~(\ref{eq:alpha-s-renorm}) and (\ref{eq:alpha-Z}).

The $\ep^0$ term of $G_1$, together with $\beta_0$ and the lowest-order 
anomalous dimension $A_1$, specify the two most singular terms $\ep^{-2n}$
and $\ep^{-2n+1}$ to all orders $\as^{\,n}$. Likewise, if (besides two more
contributions to $G_1$) also the leading term of $G_2$ and the NLO quantities
$\beta_1$ and $A_2$ are known, the resummation fixes the first four leading
poles at each order. This has been the status up to now, referred to as the
next-to-leading (NL) contributions in Section 1. 
In the next section, we will present the leading term of $G_3$ and the 
corresponding higher coefficients in the $\ep$-expansions of $G_2$ and $G_3$.
Together with $\beta_2$ (as indicated before Eq.~(\ref{eq:betaD})) and our 
recent result for $A_3$~\cite{Moch:2004pa}, these results provide the NNL terms
at all orders, especially fixing the $\ep^{-4}$ and $\ep^{-3}$ poles in 
Eq.~(\ref{eq:ff4loop}).  
\section{Fixed-order results and resummation coefficients}
\label{sec:ffactor3}
We now turn to the extraction of the quark form factor up to order $\as^{\,3}$
from our third-order computation of the deep-inelastic structure functions
\cite{Vermaseren:2005qc}. As also discussed in Refs.~\cite
{Moch:2004pa,Vogt:2004mw}, the calculation has been performed via forward
Compton amplitudes and the optical theorem. 
The cuts of the corresponding diagrams always include real-emission 
contributions, thus the purely virtual form-factor part cannot be directly 
read off at this level. 
Nevertheless we 
can reconstruct the form factor from our results, except (as explained below) 
at the highest power of $\ep$ which was consistently kept in the calculations. 
Consequently, we can derive all $1/\ep$ pole terms at order $\as^{\,3}$, since 
the forward Compton amplitudes have been computed to order $\ep^0$ for
Ref.~\cite{Vermaseren:2005qc}.

Our starting point for the determination of the form factor is the 
unrenormalized (and unfactorized) partonic structure function $F^{\rm b}$ for
$\,\gamma^{\,\ast}\! q \ra\! X\,$ in the limit $x \ra 1$, where $x$ denotes the 
partonic Bjorken variable. Using the end-point properties of the harmonic 
polylogarithms \cite{Remiddi:1999ew} in which these results are expressed, we 
remove all regular contributions and only retain the singular pieces 
proportional to $\delta(1-x)$ and the +-distributions at order $\as^{\,n}$,
\beq
  \DD_{\,k} \: = \: \left[ \frac{\ln^{\,k} (1-x)}{1-x} \right]_+ \:\: ,
  \quad\quad k \: = \: 1,\,\ldots\, 2n-1 \, .
\eeq
The resulting expressions are then compared to the general structure of the 
$n$-th order contribution $F^{\rm b}_n$ in terms of the $l$-loop form factors 
${\cal F}_l$ and the corresponding pure real-emission parts ${\cal S}_{\,l}$,
\bea
F^{\rm b}_0
     &\:=\:& \delta(1-x) \nn \\[0.5mm]
F^{\rm b}_1
     &\:=\:& 2 {\cal F}_1\,\delta(1-x) + {\cal S}_1 \nn \\[0.5mm]
F^{\rm b}_2
     &\:=\:& 2 {\cal F}_2\, \delta(1-x)  
           + \left({\cal F}_1\right)^2 \delta(1-x)
           + 2 {\cal F}_1 {\cal S}_1 + {\cal S}_2 \nn \\[0.5mm]
F^{\rm b}_3
     &\:=\:& 2 {\cal F}_3\, \delta(1-x) 
           + 2 {\cal F}_1 {\cal F}_2\, \delta(1-x)
           + 2 {\cal F}_2 {\cal S}_1 + 2 {\cal F}_1 {\cal S}_2 
           + {\cal S}_3\: .
\label{eq:Fbdec}
\eea
The $x$-dependence of the factors ${\cal S}_k$ is given by the 
$D$-dimensional +-distributions $f_{k\ep}$ defined by   
\beq
\label{eq:Dplusdist}
f_{k\epsilon}(x) \: = \: [\,(1-x)^{-1-k\epsilon}\,]_+ 
                 \: = \: - {1 \over k\epsilon}\, \delta(1-x) + \sum_{i=0}\,
                        {(-k\epsilon)^i  \over i\, !}\, \DD_{\,i} \: .
\eeq
The $\as^{\,n}$ contributions ${\cal F}_n$ and ${\cal S}_n$ in Eq.~(\ref
{eq:Fbdec}) exhibit poles in $\ep$ up to order $\ep^{-2n}$. The corresponding
bare structure function $F^{\rm b}_n$, on the other hand, only include terms
up to $\ep^{-n}$, as the higher divergences on the right-hand sides cancel for 
these inclusive quantities due to the Kinoshita-Lee-Nauenberg theorem~\cite
{Kinoshita:1962xx,Lee:1964xx}. In fact, the complete cancellation already 
occurs at the level of the individual diagrams of Ref.~\cite{Vermaseren:2005qc}
for the forward Compton amplitude.

Once the products of lower-order quantities in Eq.~(\ref{eq:Fbdec}) have been 
subtracted from $F^{\rm b}_n$, the contribution of the $n$-loop form factor
${\cal F}_n$ can be extracted by performing the substitution
\beq
  \DD_{\,0} \: \to \: {1 \over n \ep} \, \delta(1 - x)
  - \sum_{i=1}\, {(-n \ep)^i  \over i\, !}\, \DD_{\,i} 
\eeq
which eliminates, besides the +-distributions, the remaining $\delta(1-x)$ 
originating in the factor $f_{n\ep}$ of Eq.~(\ref{eq:Dplusdist}) in the
purely real part ${\cal S}_n$. However, as $\delta(1-x)$ enters $f_{n\ep}$
with a factor $1/\ep$, this extraction does not work at the highest order of 
$\ep$ kept in the calculation of $F^{\rm b}_n$. Hence, as stated above, the 
determination of the $n$-loop form factor ${\cal F}_n$ to order $\ep^k$ in this 
approach requires the calculation of the bare partonic structure function 
$F^{\rm b}_n$ to order $\ep^{k+1}$.

In addition, the subtraction of the lower-order contributions ${\cal F}_l$ and 
${\cal S}_l$ with $l<n$ in Eq.~(\ref{eq:Fbdec}) requires the extension of 
these quantities to higher orders in $\ep$. Specifically, the first- and
second-order quantities are required to order $\ep^3$ and $\ep^1$, respectively,
for the extraction of the pole terms of the three-loop form factor. These
functions have been determined from the calculation of $F^{\rm b}_1$ to order
$\ep^4$ and $F^{\rm b}_2$ to order $\ep^2$. In fact, anticipating a future
extension to the finite parts of the three-loop form factor ${\cal F}_3$, we
have extended these calculations to one more power of $\ep$, making use of
the fact that the one- and two-loop integrals for the calculation of the
structure functions were evaluated to order $\ep^5$ and $\ep^3$ anyway, see
Table 3 of Ref.~\cite{Vermaseren:2005qc}. As a check of these new two-loop 
results (the one-loop quantities are known to all orders in $\ep$ anyway), a
separate calculation of ${\cal F}_2$ has been performed to order $\ep^2$
in the approach of Refs.~\cite{vanNeerven:1985xr,Matsuura:1989sm}. The
results for the corresponding seven diagrams are listed in Appendix~B. 
                 
To the accuracy in $\ep$ just discussed, the unrenormalized quark form factor 
reads, up to three loops in the notation of Eq.~(\ref{eq:finiteorderff}),
\bea
  {\cal F}_1 & = & 
  \cf \* \biggl\{
       - 2 \* {1 \over \epsilon^2}
       - 3 \* {1 \over \epsilon}
          - 8
          + \z2
       + \epsilon  \*  \left(
          - 16
          + {3 \over 2} \* \z2
          + {14 \over 3} \* \z3
          \right)
       + \epsilon^2  \*  \left(
          - 32
          + 4 \* \z2
          + 7 \* \z3
          + {47 \over 20} \* \zs
          \right)
\nonumber\\
& &\mbox{}
       + \epsilon^3  \*  \left(
          - 64
          + 8 \* \z2
          + {56 \over 3} \* \z3
          + {141 \over 40} \* \zs
          - {7 \over 3} \* \z2 \* \z3
          + {62 \over 5} \* \z5
          \right)
       + \epsilon^4 \*  \biggl(
          - 128
          + 16 \* \z2
          + {112 \over 3} \* \z3
\nonumber\\
& &\mbox{} 
          + {47 \over 5} \* \zs
          - {7 \over 2} \* \z2 \* \z3
          + {93 \over 5} \* \z5
          + {949 \over 280} \* \zt
          - {49 \over 9} \* \zzs
          \biggr)
          \biggr\}
\label{eq:ff1loopexplicit}
\, ,  \\[3mm]
  {\cal F}_2 & = & 
  \cfs \* \biggl\{
         2 \* {1 \over \epsilon^4}
       + 6 \* {1 \over \epsilon^3}
       + {1 \over \epsilon^2} \* \left(
            {41 \over 2}
          - 2 \* \z2
          \right)
       + {1 \over \epsilon} \*  \left(
            {221 \over 4}
          - {64 \over 3} \* \z3
          \right)
       + {1151 \over 8} 
       + {17 \over 2} \* \z2 
       - 58 \* \z3 
\nonumber\\
& &\mbox{}
       - 13 \* \zs 
       + \epsilon \*  \left(
            {5741 \over 16}
          + {213 \over 4} \* \z2
          - {839 \over 3} \* \z3
          - {171 \over 5} \* \zs
          + {112 \over 3} \* \z2 \* \z3
          - {184 \over 5} \* \z5
          \right)
       + \epsilon^2 \* \biggl(
            {27911 \over 32}
\nonumber\\
& &\mbox{}
          + {1839 \over 8} \* \z2
          - {6989 \over 6} \* \z3
          - {3401 \over 20} \* \zs
          + 54 \* \z2 \* \z3
          - {462 \over 5} \* \z5
          + {223 \over 5} \* \zt
          + {2608 \over 9} \* \zzs
          \biggr)
          \biggr\}
\nonumber\\
& &\mbox{}
  + \cf \* \ca \* \biggl\{
       - {11 \over 6} \* {1 \over \epsilon^3}
       + {1 \over \epsilon^2} \* \left(
          - {83 \over 9}
          + \z2
          \right)
       + {1 \over \epsilon} \* \left(
          - {4129 \over 108}
          - {11 \over 6} \* \z2
          + 13 \* \z3
          \right)
          - {89173 \over 648}
\nonumber\\
& &\mbox{}
          - {119 \over 9} \* \z2
          + {467 \over 9} \* \z3
          + {44 \over 5} \* \zs
       + \epsilon \* \biggl(
          - {1775893 \over 3888}
          - {6505 \over 108} \* \z2
          + {6586 \over 27} \* \z3
          + {1891 \over 60} \* \zs
\nonumber\\
& &\mbox{}
          - {89 \over 3} \* \z2 \* \z3
          + 51 \* \z5
          \biggr)
       + \epsilon^2 \* \biggl(
          - {33912061 \over 23328}
          - {146197 \over 648} \* \z2
          + {159949 \over 162} \* \z3
          + {2639 \over 18} \* \zs
\nonumber\\
& &\mbox{}
          - {397 \over 9} \* \z2 \* \z3
          + {3491 \over 15} \* \z5
          - {809 \over 70} \* \zt
          - {569 \over 3} \* \zzs
          \biggr)
          \biggr\}
  + \nf \* \cf \* \biggl\{
         {1 \over 3} \* {1 \over \epsilon^3}
       + {14 \over 9} \* {1 \over \epsilon^2}
       + {1 \over \epsilon} \* \biggl(
            {353 \over 54} 
\nonumber\\
& &\mbox{}
          + {1 \over 3} \* \z2
          \biggr)
          + {7541 \over 324}
          + {14 \over 9} \* \z2
          - {26 \over 9} \* \z3
       + \epsilon \* \left(
            {150125 \over 1944}
          + {353 \over 54} \* \z2
          - {364 \over 27} \* \z3
          - {41 \over 30} \* \zs
          \right) 
\nonumber\\
& &\mbox{}
       + \epsilon^2 \* \biggl(
            {2877653 \over 11664}
          - {26 \over 9} \* \z2 \* \z3
          + {7541 \over 324} \* \z2
          - {4589 \over 81} \* \z3
          - {287 \over 45} \* \zs
          - {242 \over 15} \* \z5
          \biggr)
          \biggr\}
\label{eq:ff2loopexplicit}
\, ,  \\[3mm]
  {\cal F}_3 & = & 
  \cft \* \biggl\{
       - {4 \over 3} \* {1 \over \epsilon^6}
       - 6 \* {1 \over \epsilon^5}
       + {1 \over \epsilon^4}  \*  \left(
          - 25
          + 2 \* \z2
          \right)
       + {1 \over \epsilon^3}  \*  \left(
          - 83
          - 3 \* \z2
          + {100 \over 3} \* \z3
          \right)
       + {1 \over \epsilon^2}  \*  \biggl(
          - {515 \over 2}
\nonumber\\
& &\mbox{}
          - {77 \over 2} \* \z2
          + 138 \* \z3
          + {213 \over 10} \* \zs
          \biggr)
       + {1 \over \epsilon}  \*  \biggl(
          - {9073 \over 12}
          - {467 \over 2} \* \z2
          + {2119 \over 3} \* \z3
          + {1461 \over 20} \* \zs
          - {214 \over 3} \* \z2 \* \z3
\nonumber\\
& &\mbox{}
          + {644 \over 5} \* \z5
          \biggr)
          \biggr\}
+ \cfs \* \ca \* \biggl\{
         {11 \over 3} \* {1 \over \epsilon^5}
       + {1 \over \epsilon^4}  \*  \left(
            {431 \over 18}
          - 2 \* \z2
          \right)
       + {1 \over \epsilon^3}  \*  \left(
            {6415 \over 54}
          - {7 \over 6} \* \z2
          - 26 \* \z3
          \right)
\nonumber\\
& &\mbox{}
       + {1 \over \epsilon^2}  \*  \biggl(
            {79277 \over 162}
          + {1487 \over 36} \* \z2
          - {83 \over 5} \* \zs
          - 210 \* \z3
          \biggr)
       + {1 \over \epsilon}  \*  \biggl(
            {1773839 \over 972}
          + {38623 \over 108} \* \z2
          - {6703 \over 6} \* \z3
\nonumber\\
& &\mbox{}
          - {9839 \over 72} \* \zs
          + {215 \over 3} \* \z2 \* \z3
          - 142 \* \z5
          \biggr)
          \biggr\}
+ \cf \* \cas \* \biggl\{
       - {242 \over 81} \* {1 \over \epsilon^4}
       + {1 \over \epsilon^3}  \*  \left(
          - {6521 \over 243}
          + {88 \over 27} \* \z2
          \right)
\nonumber\\
& &\mbox{}
       + {1 \over \epsilon^2}  \*  \left(
          - {40289 \over 243}
          - {553 \over 81} \* \z2
          + {1672 \over 27} \* \z3
          - {88 \over 45} \* \zs
          \right)
       + {1 \over \epsilon}  \*  \biggl(
          - {1870564 \over 2187}
          - {68497 \over 486} \* \z2
\nonumber\\
& &\mbox{}
\nonumber\\
& &\mbox{}
          + {12106 \over 27} \* \z3
          + {802 \over 15} \* \zs
          - {88 \over 9} \* \z2 \* \z3
          - {136 \over 3} \* \z5
          \biggr)
          \biggr\}
+ \nf \* \cfs \* \biggl\{
       - {2 \over 3} \* {1 \over \epsilon^5}
       - {37 \over 9} \* {1 \over \epsilon^4}
       + {1 \over \epsilon^3}  \*  \biggl(
          - {545 \over 27}
\nonumber\\
& &\mbox{}
          - {1 \over 3} \* \z2
          \biggr)
       + {1 \over \epsilon^2}  \*  \left(
          - {6499 \over 81}
          - {133 \over 18} \* \z2
          + {146 \over 9} \* \z3
          \right)
       + {1 \over \epsilon}  \*  \biggr(
          - {138865 \over 486}
          - {2849 \over 54} \* \z2
          + {2557 \over 27} \* \z3
\nonumber\\
& &\mbox{}
          + {337 \over 36} \* \zs
          \biggl)
          \biggr\}
+ \nf \* \cf \* \ca \* \biggl\{
         {88 \over 81} \* {1 \over \epsilon^4}
       + {1 \over \epsilon^3}  \*  \left(
            {2254 \over 243}
          - {16 \over 27} \* \z2
          \right)
       + {1 \over \epsilon^2}  \*  \biggl(
            {13679 \over 243}
          + {316 \over 81} \* \z2
\nonumber\\
& &\mbox{}
          - {256 \over 27} \* \z3
          \biggr)
       + {1 \over \epsilon}  \*  \left(
            {623987 \over 2187}
          + {11027 \over 243} \* \z2
          - {6436 \over 81} \* \z3
          - {44 \over 5} \* \zs
          \right)
          \biggr\}
+ \nsq \* \cf \* \biggl\{
       - {8 \over 81} \* {1 \over \epsilon^4}
\nonumber\\
& &\mbox{}
       - {188 \over 243} \* {1 \over \epsilon^3}
       + {1 \over \epsilon^2}  \*  \left(
          - {124 \over 27}
          - {4 \over 9} \* \z2
          \right)
       + {1 \over \epsilon}  \*  \left(
          - {49900 \over 2187}
          - {94 \over 27} \* \z2
          + {136 \over 81} \* \z3
          \right)
          \biggr\}
\label{eq:ff3loopexplicit}
\, .
\eea
Here $\nf$ stands for the number of effectively massless quark flavours, $C_F$ 
and $C_A$ are the usual QCD colour factors, $C_F = 4/3$ and $C_A = 3$, and the 
values of Riemann's zeta function are denoted by $\zeta_n$.

Eq.~(\ref{eq:ff3loopexplicit}) and the $\ep^1$ and $\ep^2$ parts of 
Eq.~(\ref{eq:ff2loopexplicit}) are new results of this article. The four 
highest $1/\ep$ poles of the three-loop form factor ${\cal F}_3$ provide 
the first complete verification of the resummation of the next-to-leading 
contributions. With the anomalous dimensions (\ref{eq:abexp}) known up to 
$A_3$, the remaining two poles are sufficient to fix the NNL contributions
to the function $G$ in Eq.~(\ref{eq:Gsol}). Especially, we can derive the
first ($\ep=0$) term of the third-order function $G_3(\ep)$. 
    
Before we turn to these results we recall, for completeness, the known 
coefficients of the cusp anomalous dimension $A(\ars)$. The results for $A_1$ 
and $A_2$,
\beq
\label{eq:a12}
  A_1 \: = \: 4 \,\cf \:\: , \quad\quad
  A_2 \: = \: 8 \,\cf \ca \left( \frac{67}{18} - \z2 \right) 
              + 8 \,\cf \nf \left( - \frac{5}{9} \right) \: ,
\eeq
have been known for a long time~\cite{Kodaira:1982nh}. The recently completed
expression for $A_3$ reads~\cite{Moch:2004pa}
\bea
\label{eq:a3}
  A_3 & = &
     16\, \cf \cas \, \left( \frac{245}{24} - \frac{67}{9}\: \z2
         + \frac{11}{6}\:\z3 + \frac{11}{5}\:\zs \right)
   +  16\, \cfs \nf\, \left( -  \frac{55}{24}  + 2\:\z3 \right)
   \quad \nn \\[1mm] & & \mbox{}
   +  16\, \cf \ca \nf\, \left( - \frac{209}{108}
         + \frac{10}{9}\:\z2 - \frac{7}{3}\:\z3 \right)
   \, +\,  16\, \cf \nsq \left( - \frac{1}{27}\,\right) \: .
\eea
See Refs.~\cite{Gracey:1994nn,Moch:2002sn,Berger:2002sv} for previous partial 
results on the $\nf$-contributions. Very recently the $\zs$ term in 
Eq.~(\ref{eq:a3}) has been confirmed in Ref.~\cite{Bern:2005iz}, see the
discussion at the end of Section 4.

Inserting Eqs.~(\ref{eq:a12}) and (\ref{eq:a3}) into the resummation relations 
(\ref{eq:ff1loop}) -- (\ref{eq:ff3loop}) and comparing to the explicit results 
(\ref{eq:ff1loopexplicit}) -- (\ref{eq:ff3loopexplicit}), we obtain the 
following perturbative expansion of Eq.~(\ref{eq:Gsol}) at $\mu^2 = Q^2\,$:  
\bea
\label{eq:g1}
  G_1 & = & 
          6 \* \cf
        + \epsilon \* \cf \* (16 - 2 \* \z2)
        + \epsilon^2 \* \cf \* \left(32 - 3 \* \z2 - {28 \over 3} \* \z3\right)
        + \epsilon^3 \* \cf \* \biggl(64 - 8 \* \z2 - 14 \* \z3  
\nonumber \\
& &
          - {47 \over 10} \* \zs\biggr)
        + \epsilon^4 \* \cf \* \left(128 - 16 \* \z2 
        - {112 \over 3} \* \z3 - {141 \over 20} \* \zs 
        + {14 \over 3} \* \z2 \* \z3 - {124 \over 5} \* \z5\right)
\nonumber \\
& & + \epsilon^5 \* \cf \* \left(
          256 - 32 \* \z2 - {224 \over 3} \* \z3 - {94 \over 5} \* \zs 
        + 7 \* \z2 \* \z3 - {186 \over 5} \* \z5 - {949 \over 140} \* \zt 
        + {98 \over 9} \* \zzs 
\right)
\, ,
\\[1mm]
\label{eq:g2}
  G_2 & = & 
         \cfs \* (3 - 24 \* \z2 + 48 \* \z3)
       + \cf \* \ca \* \left({2545 \over 27} + {44 \over 3} \* \z2 
       - 52 \* \z3\right)
       + \cf \* \nf \* \left( - {418 \over 27} - {8 \over 3} \* \z2\right)
\nonumber \\
& &
        + \epsilon \* \cfs \* \left({1 \over 2} - 116 \* \z2 + 120\*\z3 
        + {176 \over 5} \* \zs\right)
        + \epsilon \* \cf \* \ca \* \left({70165 \over 162} 
        + {575 \over 9} \* \z2 - {520 \over 3} \* \z3 
\right.
\nonumber \\
& &
\left.
        - {176 \over 5} \* \zs\right)
        + \epsilon \* \cf \* \nf \* \left( - {5813 \over 81} 
        - {74 \over 9} \* \z2 + {16 \over 3} \* \z3\right)
        + \epsilon^2 \* \cfs \* \left( - {109 \over 4} - 437 \* \z2 
        + 736 \* \z3 
\right.
\nonumber \\
& &
        + {432 \over 5} \* \zs - 112 \* \z2 \* \z3 + 48 \* \z5 \biggr)
        + \epsilon^2 \* \cf \* \ca \* \biggl( {1547797 \over 972} 
        + {7297 \over 27} \* \z2 - {24958 \over 27} \* \z3 
\nonumber \\
& &
        - {653 \over 6} \* \zs + {356 \over 3} \* \z2 \* \z3 
        - 204 \* \z5 \biggr)
        + \epsilon^2 \* \nf \* \cf \* \left( - {129389 \over 486} 
        - {850 \over 27} \* \z2 
        + {1204 \over 27} \* \z3 + {7 \over 3} \* \zs \right)
\nonumber \\
& & 
        + \epsilon^3 \* \cfs  \*  \biggl(  
        - {1287 \over 8} - {2991 \over 2} \* \z2 
        + 3614 \* \z3 
        + 508 \* \zs - 104 \* \z2 \* \z3 + 72 \* \z5 
          - {6864 \over 35} \* \zt 
\nonumber \\
& & 
          - 1072 \* \zzs \biggr)
        + \epsilon^3 \* \cf \* \ca  \*  \biggl( 
          {31174909 \over 5832} 
          + {155701 \over 162} \* \z2 
          - {308810 \over 81} \* \z3 
          - {100907 \over 180} \* \zs 
\nonumber \\
& & 
          + {478 \over 3} \* \z2 \* \z3 - 840 \* \z5 
          + {1618 \over 35} \* \zt + {2276 \over 3} \* \zzs \biggr)
        + \epsilon^3 \* \nf \* \cf  \*  \biggl(   
          - {2628821 \over 2916} - {8405 \over 81} \* \z2 
\nonumber \\
& & 
          + {16340 \over 81} \* \z3
          + {1873 \over 90} \* \zs + {44 \over 3} \* \z2 \* \z3 
          + 48 \* \z5 \biggr)
\, ,
\\[1mm]
\label{eq:g3}
  G_3 & = & 
          \cft  \* \left( 29 + 36 \* \z2 + 136 \* \z3 + {576 \over 5} \* \zs 
          - 64 \* \z2 \* \z3 - 480 \* \z5 \right)
        + \cfs \* \ca  \* \left( {232 \over 3} - {2096 \over 3} \* \z2 
\right.
\nonumber \\
& &
          + {3008 \over 3} \* \z3 - {8 \over 3} \* \zs 
          + 32 \* \z2 \* \z3 + 240 \* \z5 \biggr)
        + \cf \* \cas \* \biggl( {1045955 \over 729} + {34732 \over 81} \* \z2 
          - {34928 \over 27} \* \z3 
\nonumber \\
& &
          - {188 \over 3} \* \zs 
          + {176 \over 3} \* \z2 \* \z3 + 272 \* \z5 \biggr)
        + \nf \* \cfs  \* \left(  - {3826 \over 27} + {296 \over 3} \* \z2 
          - {1232 \over 9} \* \z3 
          + {208 \over 15} \* \zs \right)
\nonumber \\
& &
        + \nf \* \cf \* \ca  \* \left(  - {309838 \over 729} 
          - {11728 \over 81} \* \z2 
          + {1448 \over 9} \* \z3 + {88 \over 15} \* \zs \right)
\nonumber \\
& &
        + \nsq \* \cf  \* \left( {19676 \over 729} + {304 \over 27} \* \z2 
          + {32 \over 27} \* \z3 \right)
\, .
\eea
As discussed at the end of Section 2, these results are sufficient to fix the
next-to-next-to-leading contributions, i.e., the six highest poles in $\ep$, 
to all orders in the strong coupling. In fact, in view of a future extension of 
$G_3$ to order $\ep$, the first- and second-order results (\ref{eq:g1}) and 
(\ref{eq:g2}) already transcend this accuracy by one power in $\ep$.

We close this section by a brief discussion of our three-loop results 
(\ref{eq:ff3loopexplicit}) and (\ref{eq:g3}). The former result for the $1/\ep$
poles of the quark form factor in massless QCD is not directly applicable to 
any physical process. For use in cross section calculations such as $e^+e^- \!
\ra 2 \mbox{ jets}$ at the next-to-next-to-next-to-leading order (N$^3$LO), one
would need the finite contribution to ${\cal F}_3$ as well. 
However, the resulting leading term (\ref{eq:g3}) of $G_3$ is of immediate 
interest for predictions of the pole structure of QCD
amplitudes at higher orders \cite{Sterman:2002qn,Kosower:2003cz} generalizing 
Catani's NNLO formula \cite{Catani:1998bh}. For the four-quark amplitude at 
$N^3$LO, ${\rm q} {\rm q} \to {\rm q} {\rm q}$, for instance, an explicit 
prediction has been derived in Ref.~\cite{Sterman:2002qn}, for which Eqs.\
(\ref{eq:ff3loopexplicit}) and (\ref{eq:g3}) now provide the last missing 
piece of information. 
\section{The time-like case and non-QCD applications}
\label{sec:ffactor4}
So far, our discussion has been restricted to space-like photon momenta, $q^2 = 
-Q^2 < 0\,$. The modifications for the time-like case $q^2 > 0$ are obtained by 
analytic continuation.  For the resummed quark form factor in Eq.~(\ref
{eq:resummedff}) this continuation has been discussed in Ref.~\cite
{Magnea:1990zb}, while the finite-order expansions (\ref{eq:finiteorderff}) 
are transferred to $q^2 > 0$ according to~\cite{vanNeerven:1985xr}
\bea
\label{eq:ancont-spacetime}
 \biggl({-q^2 \over \mu^2}\biggl)^{\! -l\epsilon} & = &
 \biggl({q^2 \over \mu^2}\biggl)^{\! -l\epsilon} \left(
 {\Gamma(1-l \epsilon) \Gamma(1+l \epsilon) \over \Gamma(1 - 2 l \epsilon) 
 \Gamma(1 + 2 l \epsilon)}
 - {\rm i} {\pi l \epsilon \over \Gamma(1 - l \epsilon) \Gamma(1 + l \epsilon)}
 \right) \, .
\eea

Of particular interest is the absolute ratio $|{\cal F}(q^2)/{\cal F}(-q^2)|$ 
of the renormalized time-like and space-like form factors. This quantity is 
infrared finite and directly enters the cross section for Drell-Yan lepton
pair production in hadronic collisions. Transforming Eqs.~(\ref{eq:ff1loop}) 
-- (\ref{eq:ff4loop}) back to the renormalized quantities using 
Eqs.~(\ref{eq:alpha-s-renorm}) and (\ref{eq:alpha-Z}), and then employing the
analytic continuation (\ref{eq:ancont-spacetime}) we obtain the expansion
\bea
\label{eq:dytmsp}
\left| {{\cal F}(q^2) \over {\cal F}(-q^2)} \right|^2  & = & 
          1 
        + \ars \* \{
            3 \* \z2 \* A_1
          \}
        + \as^{\,2} \* \biggl\{
            {9 \over 2} \* \zs \* A_1^2 
          + 3 \* \z2 \* (\beta_0 \* G_1 + A_2)
          \biggr\}
\nonumber \\
& & \mbox{}
        + \ars^{\,3} \* \biggl\{
            {9 \over 2} \* \zt \* A_1^3
          + 3 \* \zs \* A_1 \* (3 \* \beta_0 \* G_1 - \beta_0^2 + 3 \* A_2)
          + 3 \* \z2 \* (A_3 + \beta_1 \* G_1 + 2 \* \beta_0 \* G_2)
          \biggr\}
\nonumber \\
& & \mbox{}
        + \ars^{\,4} \* \biggl\{
            {27 \over 8} \* \zf \* A_1^4
          + {9 \over 2} \* \zt  \* A_1^2 \* (3 \* \beta_0 \* G_1 
          - 2 \* \beta_0^2 + 3 \* A_2)
          + {3 \over 2} \* \zs \* (
            - 6 \* \beta_0^2 \* A_2 
            + 3 \* \beta_0^2 \* G_1^2 
\nonumber \\
& & \mbox{}\quad\quad
            + 3 \* A_2^2 
            + 12 \* \beta_0 \* A_1 \* G_2 
            + 6 \* A_1 \* A_3
            + 6 \* \beta_1 \* A_1 \* G_1 
            - 5 \* \beta_0 \* \beta_1 \* A_1 
            + 6 \* \beta_0 \* A_2 \* G_1 
\nonumber \\
& & \mbox{}\quad\quad
            - 6 \* \beta_0^3 \* G_1)
          + 3 \* \z2 \* (A_4 + \beta_2 \* G_1 + 3 \* \beta_0 \* G_3 
          + 2 \* \beta_1 \* G_2)
          \biggr\} \:\: + \:\: {\cal O} (\ars^{\,5})
\eea
in terms of the couplings $\ars(q^2) = \ars(-q^2) = \ars$. Note that, since 
this ratio is infrared finite, only the $\ep = 0$ parts of the coefficients 
$G_i$ enter Eq.~(\ref{eq:dytmsp}). Consequently, all terms contributing at the 
fourth order are now known, with the exception of the four-loop cusp anomalous 
dimension $A_4$ of which only the small $n_{\! f}^{\,3}$ contribution has been 
derived so far~\cite{Gracey:1994nn}. 
 
The effect of $A_4$ is expected to be small, therefore we can nevertheless 
evaluate the ratio (\ref{eq:dytmsp}) also numerically up to the fourth order, 
employing the [1/1] Pad\'e estimate of Ref.~\cite{Moch:2005ba}, 
\beq
\label{eq:A4pade}
  A_{\rm q,4} \; \approx \; 7849\: ,\:\: 4313\: ,\:\:  1553 \quad
  \mbox{for} \quad \nf \; = \; 3\: ,\:\: 4\: ,\:\: 5 \:\: ,
\eeq
to which we assign a conservative 50\% uncertainty. Switching back to the
strong coupling $\as = 4\pi\: \ars$ at the scale $q^2$ as the expansion 
parameter, Eq.~(\ref{eq:dytmsp}) for $\nf = 4$ yields the numerical expansion
\beq
\label{eq:dytmsp-num}
\left| {{\cal F}(q^2) \over {\cal F}(-q^2)} \right|^2 \: = \: 
           1 + 2.094\: \as + 5.613\: \as^{\,2} 
             + 15.70\: \as^{\,3} + (48.63 \pm 0.43)\: \as^{\,4} \: .
\eeq
This result does not look like a nicely converging expansion, but so far does 
not exhibit a clear factorial growth of the higher-order coefficients either.
As already pointed out in Ref.~\cite{Magnea:1990zb}, the only genuine $l$-loop 
contribution at order $\as^{\,l}$ is given by the anomalous dimension $A_l$, 
which in Eq.~(\ref{eq:dytmsp-num}) contributes 24\%, 7\% and $(2\pm 1)\%$ of 
the total coefficient at the second, third and fourth order, respectively.
On the other hand, the contributions of the quantities $G_{l-1}$ at order 
$\as^{\,l}$ are large, amounting to 37\%, 41\%, 50\% at $\,l = 2,\:3,\:4$. 
Consequently, the higher-order ($\,l \geq 5\,$) terms in 
Eq.~(\ref{eq:dytmsp-num}) cannot be predicted quantitatively at this point.

Exponentiations like Eq.~(\ref{eq:resummedff}) for the form factor ${\cal F}$
have also been studied for electroweak interactions~\cite{Kuhn:2001hz}, where 
a fermion or gauge-boson mass $m$ acts as a regulator for collinear or infrared 
singularities. Of course, both the counter-term function $K$ in Eq.~(\ref
{eq:Kdeq}) and the lower integration limit in Eq.~(\ref{eq:resummedff}) are 
modified in this case, as they depend on the infrared sector of the theory.
However, the leading ($\ep =0)$ term of the function $G$ in Eq.~(\ref{eq:Gdeq})
is independent of the regulator at each order in the coupling constant.
This contribution entirely originates in the so-called hard region in an
expansion of the loop integrals in different regions~\cite{Smirnov:2002pj}. 
In this region all loop momenta are of order $Q$, effectively leading to 
the massless case considered in Eqs.~(\ref{eq:ff3loop}) and 
(\ref{eq:ff3loopexplicit}).
 
This `universality' implies, for instance, that Eq.~(\ref{eq:g3}) provides a 
prediction for the coefficient of $\,\ln(Q^2/m^2)\,$ in the three-loop
quantity ${\cal F}_3$ for an Abelian gauge theory with fermion masses like 
Quantum Electrodynamics (QED) \cite{Penin:2005kf} after the usual 
identification of the colour factors. For QED, e.g., one has $\,C_F=1$, $C_A=0$
and $T_f=1$ instead of our QCD convention $T_f\, \nf =\nf/2$.

Another interesting implication of Eq.~(\ref{eq:ff3loopexplicit}) arises for 
maximally supersymmetric Yang-Mills theory (MSYM), i.e., Yang-Mills theory with 
${\cal N}=4$ supersymmetry in four dimensions. QCD results may be carried over 
to this theory using the inspired observation~\cite{Kotikov:2004er} that the 
MSYM results can be obtained from the contributions of leading transcendentality
in QCD. This procedure has been applied to the QCD results for the three-loop 
anomalous dimensions of spin $N$ of leading-twist operators
\cite{Moch:2004pa,Vogt:2004mw}, which were employed to extract corresponding
quantities in MSYM~\cite{Kotikov:2004er}. 
Strikingly enough, the resulting MSYM anomalous dimensions completely agree 
with predictions based on integrability for the planar three-loop contribution 
to the dilatation operator~\cite{Beisert:2003tq}. 
This agreement has been checked up to spin $N=8$ in Ref.~\cite
{Staudacher:2004tk} and is now established up to $N=70$~\cite
{Staudacher:privatecomm} (for a review see also Ref.~\cite{Beisert:2004ry}).
 
Although no formal proof exists for the procedure of Ref.~\cite{Kotikov:2004er},
it has recently been used in reverse, namely to predict terms of highest 
transcendentality in the QCD form factor.
Based on studies of planar amplitudes in MSYM at three loops~\cite{Bern:2005iz},
where an interesting pattern of iteration for the four-point amplitude has been
found, both the coefficients $A_l \bigl|_{\rm MSYM}$ and the leading 
contribution to $G_l \bigl|_{\rm MSYM}$ have been determined for $l \leq 3$.
Our new result for the three-loop form factor ${\cal F}_3$ in 
Eq.~(\ref{eq:ff3loopexplicit}) and for coefficient $G_3$ in Eq.~(\ref{eq:g3}) 
puts us in a position to check this part of Ref.~\cite{Bern:2005iz} and thereby
provide further evidence on the procedure of Ref.~\cite{Kotikov:2004er}.

The only transcendental numbers entering the results for the form factor are
the values $\zeta_n$ of Riemann's zeta function. Hence the procedure of 
Ref.~\cite{Kotikov:2004er} implies that, as each order in $\as$, one keeps only
the highest terms $\zeta_n$ and $\zeta_i\,\zeta_j$ with $\,i+j=n$.
After the SYM identification $C_A = C_F = n_c$ (terms with $\nf$ do not 
contribute at the highest transcendentality), Eqs.~(\ref{eq:a12}) and 
(\ref{eq:a3}) lead to 
\bea
\label{eq:a13susy}
  A_1 \Bigl|_{\rm MSYM}  \: = \: 
         4\, \* n_c
\:\: , \quad\quad
  A_2 \Bigl|_{\rm MSYM}  \: = \: 
         - 8\, \* \z2 \* n_c^2
\:\: , \quad\quad
  A_3 \Bigl|_{\rm MSYM}  \: = \: 
         {176 \over 5}\, \* \zs \* n_c^3
\, .
\eea
Correspondingly, Eqs.~(\ref{eq:g1})--(\ref{eq:g3}) result in
\bea
\label{eq:g13susy}
  G_1 \Bigl|_{\rm MSYM}  \: = \: 
0
\: , \quad\quad
  G_2 \Bigl|_{\rm MSYM}  \: = \: 
       - 4\, \* \z3 \* n_c^2
\: , \quad\quad
  G_3 \Bigl|_{\rm MSYM}  \: = \: 
         {80 \over 3}\, \* \z2 \* \z3 \* n_c^3
       + 32\, \* \z5 \* n_c^3
\, . 
\eea
Both relations agree with the results of Ref.~\cite{Bern:2005iz}, and hence 
with the prescription of Ref.~\cite{Kotikov:2004er}. 
\section{Summary}
\label{sec:sum}
We have derived new higher-order QCD results for the electromagnetic form 
factor of on-shell massless quarks. Specifically, we have extracted all 
third-order $1/\ep$ pole terms in dimensional regularization from our recent
computation of the three-loop coefficient functions for inclusive 
deep-inelastic scattering~\cite{Vermaseren:2005qc}, supplemented by a 
higher-$\ep$ extension of the two-loop contributions. These results, together 
with our extension of the resummation of the form factor to the next-to-next-%
to-leading contributions, fix the six highest $1/\ep$ poles to all orders.
As an example, we have provided the explicit expression for the coefficients
of $\ep^{-8}\,\ldots\,\ep^{-3}$ at four loops.

While the pole terms of the form factor alone are not sufficient for use in
other three-loop calculations like $e^+e^- \!\ra 2 \mbox{ jets}$, they do have
immediate theoretical applications both for the infrared structure of 
higher-order QCD amplitudes and for other gauge theories such as QED and 
$\,{\cal N}\! = 4\,$ Super-Yang-Mills theory, where our results confirm a 
recent corresponding calculation in Ref.~\cite{Bern:2005iz}. 
Moreover, our present results are sufficient (up to a numerically irrelevant 
uncertainty due to the unknown four-loop cusp anomalous dimension) for 
extending the finite absolute ratio of the time-like and space-like form 
factors, which directly enters the description of the Drell-Yan process, to the
fourth order in $\as$.
  
We close by noting that the computation of the finite part of the three-loop 
quark form factor ${\cal F}_3$ by an extension of the techniques employed in 
this article is feasible. 
\section*{Acknowledgments}
We thank W.~Giele, N.~Glover, E.~Laenen, A.~Penin and P.~Uwer for stimulating 
discussions. The Feynman diagrams in Appendix B have been drawn with the 
packages {\sc Axodraw} \cite{Vermaseren:1994je} and {\sc Jaxo\-draw} 
\cite{Binosi:2003yf}.
The work of S.M. has been supported in part by the Helmholtz Gemeinschaft 
under contract VH-NG-105 and by the Deutsche Forschungsgemeinschaft in 
Sonderforschungs\-be\-reich/Transregio 9.
The work of J.V. has been part of the research program of the Dutch Foundation 
for Fundamental Research of Matter (FOM).
\renewcommand{\theequation}{A.\arabic{equation}}
\setcounter{equation}{0}
\section*{Appendix A}
Here we give some details for the determination of the loop-expanded form 
factor from Eq.~(\ref{eq:resummedff}). Useful auxiliary relations are 
\bea
\label{eq:1mxexp}
{1 \over (1-x)^{n-\epsilon}} &\: = \: & 
\sum\limits_{i=0}^\infty\, {\Gamma(n-\epsilon+i) \over \Gamma(n-\epsilon)}\, 
{x^{\,i} \over i!}\; , 
\\[1mm]
{\ln^k(1-x) \over (1-x)^{n-\epsilon}} &\:  =\: &
\left( {\partial\over \partial \epsilon} \right)^k {1 \over (1-x)^{n-\epsilon}}
\; ,
\eea
where Eq.~(\ref{eq:1mxexp}) holds for $|x| < 1$. 
The expansion of the Gamma function in powers of $\epsilon$ for positive 
integers $n$ reads
\bea
\label{eq:expgamma}
{ \Gamma(n+1+\epsilon) \over  n!\, \Gamma(1+\epsilon)} &=& 
         1
       + \epsilon \* 
            S_{1}(n)
       + \epsilon^2 \* \bigl(
            S_{1,1}(n)
          - S_{2}(n)
          \bigr)
       + \epsilon^3 \* \bigl(
            S_{1,1,1}(n)
          - S_{1,2}(n)
          \bigr)
\nonumber\\[-1mm]
& &\mbox{}
          - S_{2,1}(n)
          + S_{3}(n)
       + \epsilon^4 \* \bigl(
            S_{1,1,1,1}(n)
          - S_{1,1,2}(n)
          - S_{1,2,1}(n)
\nonumber\\[0.5mm]
& &\mbox{}
          + S_{1,3}(n)
          - S_{2,1,1}(n)
          + S_{2,2}(n)
          + S_{3,1}(n)
          - S_{4}(n)
          \bigr)
+ {\cal O}(\epsilon^5) \, ,
\eea
with $S_{m_1,\dots,m_k}(N)$ denoting the harmonic sums~\cite{Vermaseren:1998uu}.
Finally integrals of the following types occur:
\bea
\label{eq:auxint1}
\int {d \lambda \over \lambda}\:  \lambda^{-n\epsilon} &\: = \: & 
- {1 \over n \epsilon} \lambda^{-n\epsilon}\, , \\[2mm]
\label{eq:auxint2}
\int {d \lambda \over \lambda} \left(\lambda^{-\epsilon}-1\right)^n &\: = \: & 
- {1 \over \epsilon} (-1)^n \sum\limits_{j=1}^n {(-1)^j \over j} 
\left(\lambda^{-\epsilon}-1\right)^{j} 
\nonumber\\ & &
- {1 \over \epsilon} (-1)^n \sum\limits_{j=1}^n 
\left(
\begin{array}{c}
n \\
j
\end{array}
\right)
{(-1)^j \over j} 
+ (-1)^n \ln \,\lambda
\, ,\quad\quad \\[2mm]
\label{eq:auxint3}
\int {d \lambda \over \lambda}\: \lambda^{-\epsilon}
\left(\lambda^{-\epsilon}-1\right)^n &=& 
- {1 \over \epsilon} {1 \over n + 1} \left(\lambda^{-\epsilon}-1\right)^{n+1} 
\, , \\[2mm]
\label{eq:auxint4}
\int {d \lambda \over \lambda}\: \lambda^{-2\epsilon}
\left(\lambda^{-\epsilon}-1\right)^n &=& 
- {1 \over \epsilon} \left( {1 \over n + 1} + \lambda^{-\epsilon} \right) 
{1 \over n + 2} \left(\lambda^{-\epsilon}-1\right)^{n+1} \, ,
\eea
and so on, where $n>0$.
The integration over $\lambda$ and $\xi$ in Eq.~(\ref{eq:resummedff}) leads to 
double sums which are readily evaluated to any finite order in $\as$. Also 
all-order analytical results for the exponent, cf.~Ref.~\cite{Magnea:2000ss}, 
can be obtained along these lines by employing the algorithms for the 
evaluation of nested sums~\cite{Moch:2001zr}.
\renewcommand{\theequation}{B.\arabic{equation}}
\setcounter{equation}{0}
\section*{Appendix B}
\begin{figure}[htb]
  \centering
  \vspace*{5mm}
  \includegraphics[width=15.0cm]{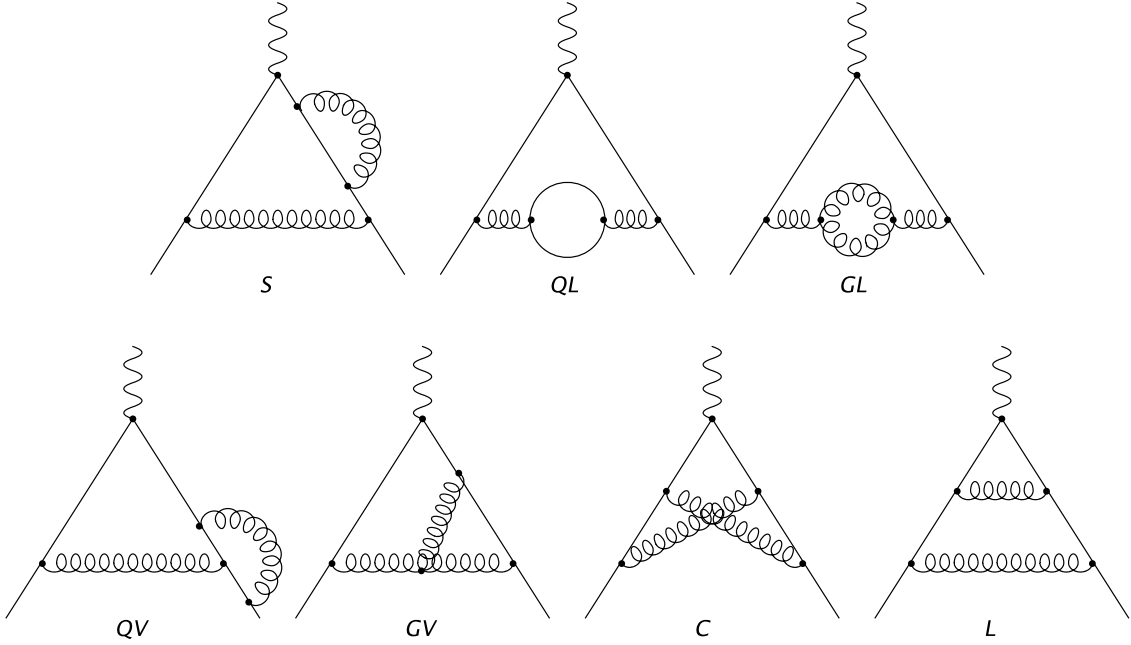}
  \caption{The Feynman diagrams contributing to the two-loop quark 
  form factor in the notation of Ref.~\cite{Matsuura:1989sm}.}
  \vspace*{5mm}
  \label{fig:ff2diags}
\end{figure}

Finally we present the results for the individual Feynman diagrams, displayed in 
Fig~\ref{fig:ff2diags}, which contribute to the two-loop form factor in the
approach (and notation) of Refs.~\cite{vanNeerven:1985xr,Matsuura:1989sm}.
The diagrams add up to the bare quark form factor in 
Eq.~(\ref{eq:ff2loopexplicit}) according to
\bea
  {\cal F}_2 &\: =\: & 2 S + QL + GL + 2 QV + 2 GV + C + L
\, .
\eea
Note that both the normalization in Eq.~(\ref{eq:finiteorderff}), where we have 
pulled out the factor $({Q^2/\mu^2})^{-2\epsilon}$, and our convention for 
$\ep$ are different from those in Ref.~\cite{Matsuura:1989sm}.

\noindent
The results for the individual diagrams to order $\epsilon^2$ are given by
\bea
  S & = & 
  \cfs  \*  \biggl\{
         {1 \over \epsilon^3} 
       + {7 \over 2} \* {1 \over \epsilon^2}           
       + {1 \over \epsilon}  \*  \biggl(
            {53 \over 4}
          - \z2
          \biggr)
          + {355 \over 8}
          - {7 \over 2} \* \z2
          - {32 \over 3} \* \z3
       + \epsilon  \*  \biggl(
            {2281 \over 16}
          - {53 \over 4} \* \z2
          - {112 \over 3} \* \z3
          - {57 \over 10} \* \zs
          \biggr)
\nonumber \\
& & \mbox{}
       + \epsilon^2  \* \biggl(
            {14299 \over 32}
          - {355 \over 8} \* \z2
          - {424 \over 3} \* \z3
          - {399 \over 20} \* \zs
          + {32 \over 3} \* \z2 \* \z3
          - {272 \over 5} \* \z5
          \biggr)
          \biggr\}
\label{eq:diaS}
\, ,\\[2mm]
  QL & = & 
  \cf \* \nf  \*  \biggl\{
         {1 \over 3} \* {1 \over \epsilon^3}            
       + {14 \over 9} \* {1 \over \epsilon^2} 
       + {1 \over \epsilon} \* \biggl(
            {353 \over 54}
          + {1 \over 3} \* \z2
          \biggr)
          + {7541 \over 324}
          + {14 \over 9} \* \z2
          - {26 \over 9} \* \z3
       + \epsilon \* \biggl(
            {150125 \over 1944}
          + {353 \over 54} \* \z2
\nonumber \\
& & \mbox{}
          - {364 \over 27} \* \z3
          - {41 \over 30} \* \zs
          \biggr)
       + \epsilon^2 \* \biggl(
            {2877653 \over 11664}
          + {7541 \over 324} \* \z2
          - {4589 \over 81} \* \z3
          - {287 \over 45} \* \zs
          - {26 \over 9} \* \z2 \* \z3
          - {242 \over 15} \* \z5
          \biggr)
          \biggr\}
\, ,
\nonumber \\
& & \mbox{}
\label{eq:diaQL}
\\[2mm]
  GL & = & 
  \cf \* \ca  \*  \biggl\{
       - {5 \over 6} \* {1 \over \epsilon^3}
       - {38 \over 9} \* {1 \over \epsilon^2}
       + {1 \over \epsilon} \* \biggl(
          - {1969 \over 108}
          - {5 \over 6} \* \z2
          \biggr)
          - {43165 \over 648}
          - {38 \over 9} \* \z2
          + {65 \over 9} \* \z3
       + \epsilon \* \biggl(
          - {873877 \over 3888}
\nonumber \\
& & \mbox{}
          - {1969 \over 108} \* \z2
          + {988 \over 27} \* \z3
          + {41 \over 12} \* \zs
          \biggr)
       + \epsilon^2 \* \biggl(
          - {16929277 \over 23328}
          - {43165 \over 648} \* \z2
          + {25597 \over 162} \* \z3
          + {779 \over 45} \* \zs
\nonumber \\
& & \mbox{}
          + {65 \over 9} \* \z2 \* \z3
          + {121 \over 3} \* \z5
          \biggr)
          \biggr\}
\label{eq:diaGL}
\, ,\\[2mm]
  QV & = & 
  \cf \* \bigl(\cf-{\ca \over 2}\bigr)  \*  \biggl\{
       - {1 \over \epsilon^3}
       - {1 \over \epsilon^2} \* \biggl(
            {11 \over 2}
          - 2 \* \z2
          \biggr)
       + {1 \over \epsilon} \* \biggl(
          - {109 \over 4}
          + 10 \* \z2
          + 2 \* \z3
          \biggr)
          - {911 \over 8}
          + {91 \over 2} \* \z2
          + {59 \over 3} \* \z3
\nonumber \\
& & \mbox{}
          + {8 \over 5} \* \zs
       + \epsilon \* \biggl(
          - {6957 \over 16}
          + {689 \over 4} \* \z2
          + {296 \over 3} \* \z3
          + {129 \over 10} \* \zs
          - {58 \over 3} \* \z2 \* \z3
          + 6 \* \z5
          \biggr)
       + \epsilon^2 \* \biggl(
          - {49639 \over 32}
\nonumber \\
& & \mbox{}
          + {4843 \over 8} \* \z2
          + {1307 \over 3} \* \z3
          + {1267 \over 20} \* \zs
          - {293 \over 3} \* \z2 \* \z3
          + {407 \over 5} \* \z5
          - {281 \over 35} \* \zt
          - {58 \over 3} \* \zzs
          \biggr)
          \biggr\}
\label{eq:diaQV}
\, ,\\[2mm]
  GV & = & 
  \cf \* \ca  \*  \biggl\{
         {1 \over 4} \* {1 \over \epsilon^4}
       + {1 \over \epsilon^2} \* \biggl(
          - {5 \over 4}
          - {1 \over 4} \* \z2
          \biggr)
       + {1 \over \epsilon} \* \biggl(
          - {73 \over 8}
          + {1 \over 2} \* \z2
          - {8 \over 3} \* \z3
          \biggr)
          - {663 \over 16}
          + {15 \over 4} \* \z2
          + {1 \over 2} \* \z3
          - {57 \over 40} \* \zs
\nonumber \\
& & \mbox{}
       + \epsilon \* \biggl(
          - {5093 \over 32}
          + {159 \over 8} \* \z2
          + {95 \over 6} \* \z3
          + {2 \over 5} \* \zs
          + {8 \over 3} \* \z2 \* \z3
          - {68 \over 5} \* \z5
          \biggr)
       + \epsilon^2 \* \biggl(
          - {35887 \over 64}
          + {1289 \over 16} \* \z2
\nonumber \\
& & \mbox{}
          + {1297 \over 12} \* \z3
          + {73 \over 8} \* \zs
          - {29 \over 6} \* \z2 \* \z3
          + {3 \over 2} \* \z5
          - {1343 \over 280} \* \zt
          + {128 \over 9} \* \zzs
          \biggr)
          \biggr\}
\label{eq:diaGV}
\, ,\\[2mm]
  C & = & 
  \cf \* \bigl(\cf-{\ca \over 2}\bigr)  \*  \biggl\{
         {1 \over \epsilon^4}
       + {4 \over \epsilon^3} 
       + {1 \over \epsilon^2} \* (
            16
          - 7 \* \z2
          )
       + {1 \over \epsilon} \* \biggl(
            58
          - 16 \* \z2
          - {122 \over 3} \* \z3
          \biggr)
          + 204
          - 58 \* \z2
          - {380 \over 3} \* \z3
\nonumber \\
& & \mbox{}
          - {53 \over 2} \* \zs
       + \epsilon \* \biggl(
            697
          - 181 \* \z2
          - {1646 \over 3} \* \z3
          - {402 \over 5} \* \zs
          + {326 \over 3} \* \z2 \* \z3
          - {842 \over 5} \* \z5
          \biggr)
       + \epsilon^2 \* \biggl(
            {4631 \over 2}
\nonumber \\
& & \mbox{}
          - {1141 \over 2} \* \z2
          - {6293 \over 3} \* \z3
          - {1744 \over 5} \* \zs
          + {836 \over 3} \* \z2 \* \z3
          - {2708 \over 5} \* \z5
          + {1399 \over 70} \* \zt
          + {4274 \over 9} \* \zzs
          \biggr)
          \biggr\}
\label{eq:diaC}
\, ,\\[2mm]
  L & = & 
  \cfs  \*  \biggl\{
         {1 \over \epsilon^4}
       + {2 \over \epsilon^3}
       + {1 \over \epsilon^2} \* \biggl(
            {17 \over 2}
          + \z2
          \biggr)
       + {1 \over \epsilon} \* \biggl(
            {101 \over 4}
          - 2 \* \z2
          + {46 \over 3} \* \z3
          \biggr)
          + {631 \over 8}
          - {35 \over 2} \* \z2
          + {152 \over 3} \* \z3
          + {103 \over 10} \* \zs
\nonumber \\[2mm]
& & \mbox{}
       + \epsilon \* \biggl(
            {3941 \over 16}
          - {335 \over 4} \* \z2
          + {439 \over 3} \* \z3
          + {159 \over 5} \* \zs
          - {98 \over 3} \* \z2 \* \z3
          + {598 \over 5} \* \z5
          \biggr)
       + \epsilon^2 \* \biggl(
            {24495 \over 32}
          - {2573 \over 8} \* \z2
\nonumber \\
& & \mbox{}
          + {2065 \over 6} \* \z3
          + {1839 \over 20} \* \zs
          - {152 \over 3} \* \z2 \* \z3
          + {1976 \over 5} \* \z5
          + {2847 \over 70} \* \zt
          - {1318 \over 9} \* \zzs
          \biggr)
          \biggr\}
\label{eq:diaL}
\, .
\eea
The loop integrations have been reduced with integration-by-parts identities 
\cite{'tHooft:1972fi,Chetyrkin:1981qh} to so-called master integrals.
This step has been automatized in Ref.~\cite{Anastasiou:2004vj}.
All master integrals except the basic non-planar triangle can be expressed in
terms of Gamma functions, thus they can be readily expanded to any order in 
$\epsilon$.
The non-planar triangle (see, e.g., Ref.~\cite{Smirnov:2004ym}) can be written 
as a double sum over Gamma functions. 
After expansion, the sums can be solved in terms of the Riemann zeta function
to any order in $\epsilon$ using the algorithms for harmonic sums
\cite{Vermaseren:1998uu} coded, as all our symbolic manipulations, in 
{\sc Form}~\cite{Vermaseren:2000nd}.

{\small


\providecommand{\href}[2]{#2}\begingroup\raggedright\endgroup

}

\end{document}